\documentstyle[psfig,lprocl,11pt]{article}
\def\Journal#1#2#3#4{{#1} {\bf #2}, #3 (#4)}

\def\PR{\em Phys. Rev.}
\def\PRL{\em Phys. Rev. Lett.}
\def\PRB{{\em Phys. Rev.} B}
\def\PRE{{\em Phys. Rev.} E}
\def\ZPB{{\em Z. Phys.} B}
\def\JPC{{\em J. Phys.} C}
\def\JPA{{\em J. Phys.} A}
\def\JMMM{\em J. Mag. Mag. Mater.}
\def\JAP{\em J. Appl. Phys.}
\def\EPL{\em Europhys.\ Lett.}
\def\SSC{\em Sol. St. Comm.}

\begin{document}
\title{EXPERIMENTS ON THE RANDOM FIELD ISING MODEL}
\author{ D.P. BELANGER }
\address{Department of Physics, University of California\\
Santa Cruz, CA 95064, USA}
\maketitle\abstracts{
New advances in experiments on the random-field Ising model,
as realized in dilute antiferromagnets,
have brought us much closer to a full characterization of the
static and dynamic critical behavior of the
unusual phase transition in three dimensions ($d=3$).
The most important experiments that have laid the ground work for our
present understanding are reviewed.  Comparisons of the data
with Monte Carlo simulations of the $d=3$ critical behavior
are made.  We review the current experimental understanding of
the destroyed $d=2$ transition and the experiments
exploring the $d=2$ metastability at low $T$.
Connections to theories most relevant to the interpretations of all the
experiments are discussed.
}

\section{Introduction}

The random-field Ising model~\cite{im75} (RFIM)
has been an important focus of theoretical and experimental
studies of the statistical physics of random and
frustrated systems.  Although there are some similarities,
particularly at large random fields,
to the physics of spin-glasses~\cite{my92}, also covered in this book,
the three dimensional ($d=3$) ground state of the RFIM
in the small random-field limit has
the same long-range order as would be observed in the absence
of random fields.  Hence, the two models differ fundamentally.
Nevertheless, the $d=3$ RFIM transition is profoundly
altered by the random field.  For $d=2$ the random field
destroys the transition which takes place in the absence
of the random field.  Not only does the RFIM have significance in
the formation of long-range order in real materials, where
defects causing random fields are often present, it also
challenges the methods and ideas of theorists and experimentalists
that have been developed in past studies of phase transitions in
pure, translationally invariant materials.  There are a
number of relevant reviews that have
been written covering the formidable problems
encountered in the experimental study of RFIM systems~\cite{by92,b92}.
This one represents a comprehensive overview of the
experimental situation in the most
studied systems, the dilute anisotropic antiferromagnets, emphasizing
the most current experimental results.
The theories and computer simulations most relevant to interpretations
of the behavior observed in dilute antiferromagnets will be included.
A few systems that are not antiferromagnets will be mentioned
in section 11.
A comprehensive review of the theory of the RFIM
by Nattermann also appears in this book.

For $d=3$ it has been rigorously~\cite{bk87}
shown that a transition must take place for small random fields.
As we shall see, the RFIM transition is very different from the more usual
phase transitions encountered in antiferromagnets.
The RFIM can be most simply modeled by spins on a lattice that
point along one axis and are subjected to
a random ordering field that competes with the long-range
collective spin ordering.
One simple Hamiltonian representing an
Ising ferromagnet with an imposed random field is
\begin{equation}
{\cal H} = - \sum_{<i,j>} J_{ij} S_i S_j - \sum_{i} h_i S_i \quad .
\label{ham}
\end{equation}
The random field has the properties
$[ h_i ]_{av} = 0$ and $[h_i^2 ]_{av} = h_r^2$
where $[ ... ]_{av}$ denotes an average over the disorder.
Most of the theoretical and simulation
efforts, though not all, have focused on such ferromagnetic
models.  On the other hand, the most studied and best
characterized experimental realization of the RFIM, by far, is the
dilute, anisotropic antiferromagnet in a uniform field applied along
the spin ordering axis, which can be represented by the Hamiltonian
\begin{equation}
{\cal H} = \sum_{<i,j>} J_{ij} \epsilon_i \epsilon_j
S_i S_j - \sum_{i} H \epsilon_i S_i \quad ,
\label{afham}
\end{equation}
where $\epsilon_i = 1$ if site $i$ is occupied and 0 if empty, and $H$ is 
the uniform field.  Locally, the sublattice with the most spins
tends to align with the applied field in competition with
long-range antiferromagnetic order in which one sublattice
globally aligns with the field.  The applied uniform field
and the effective random field generated by it are proportional~\cite{c84}.
The random field is therefore easily controlled or even
turned off completely.  This provides
the opportunity to do scaling studies not easily done in other
systems.  Importantly, samples can be cooled in zero field before
applying the random field (ZFC).  Other systems, such as those
with structural phase transitions can only be cooled in
the random field (FC). Since, as we shall see, hysteresis
plays an important role in the understanding of the RFIM transition,
the ZFC process is crucial.
Of course, by virtue of critical behavior universality,
the systems studied need not correspond precisely to the Hamiltonians
above but must simply have the appropriate symmetries.

Fishman and Aharony~\cite{fa79} first noted that the dilute antiferromagnet
in a uniform field is a RFIM system and Cardy~\cite{c84} showed
that the critical behavior in the limit of small fields
belongs to the same universality
class as the uniform ferromagnet with random fields.
These works opened up a tremendous opportunity to
investigate the RFIM experimentally.  
An understanding of the RFIM phenomena in
the dilute antiferromagnet is steadily evolving with
experiments performed on very high quality anisotropic
crystals.  A major aim of this review is to present an
overview of the $d=3$ RFIM transition that
takes place in dilute antiferromagnetic systems which
is consistent with all of the published data (though certainly
not all the published interpretations of the data).
The $d=3$ phase diagram has proven much richer than anticipated
and this review necessarily encompasses high, intermediate and low
magnetic concentrations as well as large and small
random fields.  The most recent experiments by Slani\v{c}, et al.\ \cite{sbf96}
at high magnetic concentrations are promising as they
appear to afford the opportunity to make real headway
in the experimental characterization of the RFIM critical behavior and
in making connections to recent theoretical and simulation results.
Such work is still in progress, so only preliminary
results can be discussed.

Theory and experiments on the RFIM
have been closely tied throughout the period
of investigation from the Fishman and Aharony~\cite{fa79}
work until the present, though there has not
always been agreement.  The greatest progress in the
experimental investigations has come when a variety
of techniques are employed and interpretations consistent
with them are made.  Often mistakes have been made
when only one technique is relied upon for interpretation.
A complication of studies using the dilute antiferromagnet
is that random magnetic vacancies constitute strong
pinning sites for domain walls~\cite{nv88}.  Such strong vacancy pinning,
while enriching the $d=3$ antiferromagnetic phase diagram,
is not present in the theoretically well-studied ferromagnetic model.
Random-field pinning, present in both antiferromagnets and ferromagnets,
seems to be much weaker.
The correspondence between antiferromagnets and
the ferromagnetic models is best when the magnetic dilution
is small, in which case the antiferromagnetic order is
stable up to the transition, $T_c(H)$.  For concentrations
near $x=0.5$, there is evidence that the long-range order
breaks into static structure consisting of large, intertwined
and weakly interacting domains well below $T_c(H)$.
This has prevented, at these concentrations, a characterization
of ${M_s}^2$ vs.\ $T$ and the line shape below $T_c(H)$.
When the percolation threshold is approached ($x \approx x_p = 0.25$) a de
Almeida-Thouless~\cite{at78} behavior appears for larger $H$ and the system
appears to behave similarly to a spin-glass.
In this review each of these three concentration regions is discussed.

For the case of $d=2$ dilute antiferromagnets, the random-exchange Ising model
(REIM) transition is expected to be destroyed~\cite{im75,b84} as soon as
$H$, which generates the random field, is applied,
and this has been observed~\cite{fkjcg83,i83}.
The temperature regime well below the rounded transition, however,
is still being investigated theoretically~\cite{neu96}
and experimentally~\cite{bjkb96}.
Both temperature regimes are briefly reviewed.

\section{Sample Considerations}

The most studied dilute $d=3$ antiferromagnet suitable
for RFIM studies is $Fe_xZn_{1-x}F_2$.  Its
large crystal-field anisotropy persists~\cite{hrg70}
as the magnetic spins are diluted
and it is therefore an excellent Ising system
for all ranges of magnetic concentration $x$.
Crystals can be grown for all $x$ with
extremely small concentration variations $\delta x < 10^{-3}$
and with superb structural quality.
The magnetic interactions are dominated by the second-nearest- neighbor
super-exchange between the body-center
and body-corner ions.  All other interactions
are negligible, except possibly near the percolation
threshold concentration, where even tiny frustrating interactions
become important~\cite{syp79,rcm95}.  Another class of materials representing
the anisotropic random-field systems is
$Fe_xMg_{1-x}Cl_2$.  This system differs from
$Fe_xZn_{1-x}F_2$ in that it is layered.  The layers
are ferromagnetic and the interplanar antiferromagnetic
coupling is comparable in strength
to the intraplane coupling, making this a good
$d=3$ Ising system.  The smaller exchange in this system
allows the large field region of the phase diagram to be
explored~\cite{mkbfk94}.
For $x<0.55$, a strong second-nearest-neighbor
competing exchange in the $Fe_xMg_{1-x}Cl_2$ system induces
spin-glass behavior~\cite{wmpmysi85} and so random-field studies are
restricted to higher $x$.
There is excellent agreement between
the random-field behavior of $Fe_xMg_{1-x}Cl_2$
and that of $Fe_xZn_{1-x}F_2$.
Some studies have also made use of the highly anisotropic
$Co_xZn_{1-x}F_2$ system.  A number of studies have
been made in the less anisotropic system
$Mn_xZn_{1-x}F_2$.  The anisotropy in $Mn_xZn_{1-x}F_2$, which is small
for $x=1.0$, decreases further upon dilution.  Nevertheless,
the $H=0$ REIM critical behavior
of $Mn_xZn_{1-x}F_2$~\cite{mcybub?} is quite consistent with that of
$Fe_xZn_{1-x}F_2$~\cite{bkj86} and
all of the RFIM experiments done on $Mn_xZn_{1-x}F_2$ seem
qualitatively consistent with those done in $Fe_xZn_{1-x}F_2$
and $Fe_xMg_{1-x}Cl_2$.  The system does allow large
applied fields relative to the anisotropy, allowing studies of
the spin-flop region~\cite{sof84,s82}.
For the $d=2$ RFIM, $Rb_2Co_xMg_{1-x}F_4$
has been studied and appears to be an ideal
system~\cite{ih78}.  It is very anisotropic and consists
of layers of magnetic ions with a single dominant
intralayer exchange interaction and an interlayer
interaction which is smaller by several orders of magnitude.

Disagreements among the various interpretations of experimental data
have arisen when concentration gradients obscured the
true random-field behavior of a sample and were not
fully appreciated in the data analyses.  Although
the gradient effects have been extensively reviewed~\cite{kfjb88,rkj88},
the problem is still relevant to interpretations
of recent experiments, as discussed below.
Basically, one must realize that a concentration
gradient will round a transition and can affect critical
behavior measurements drastically.  It is best if the
gradients are unambiguously determined independently of the critical
behavior measuring techniques.  The size of the gradient
can then be incorporated into the interpretation
of the critical behavior data.  Disagreements over interpretations
of data in RFIM systems are usually
resolved once the effect of concentration
gradients are properly taken into account.

\section{Scaling Behavior Theory}

Although the scaling behavior of the RFIM has been discussed
extensively in previous reviews, we emphasize the salient points
again since many experiments are addressing the RFIM critical
behavior and, unfortunately, not all of the current
experimental interpretations being proposed are consistent
with scaling theory.
Static critical behavior for temperatures very
close to the second-order transition temperature $T_c$
can generally be described by power law behaviors
which become exact as the reduced temperature $t=T/T_c-1 \rightarrow 0$.
We briefly list the ones most useful to us.  The
free energy has the asymptotic behavior
$F \sim |t|^{2- \alpha } $,
and the specific heat is correspondingly given by
\begin{equation}
C_p = A ^ { \pm } |t| ^ {- \alpha } + B \quad ,
\label{specheat}
\end{equation}
where we include a constant background term
which describes the peak height when $\alpha < 0$.
For the case where $\alpha \rightarrow 0 $ and
$A^+/A^- \rightarrow 1$, we use
the symmetric logarithmic form
\begin{equation}
C_p = A \ln|t| + B
\label{spec-heat}
\end{equation}
instead.  Several critical parameters can be obtained from
neutron scattering~\cite{c89}.  The
correlation length for antiferromagnetic fluctuations has the form
\begin{equation}
\xi = \kappa ^{-1} = \xi _o ^ { \pm } |t| ^ {- \nu } \quad .
\label{xi}
\end{equation}
The staggered susceptibility is
\begin{equation}
\chi _s = \chi _o ^ { \pm } |t| ^ {- \gamma } \quad .
\label{chi}
\end{equation}
For random-field systems we have the disconnected
staggered susceptibility
\begin{equation}
{\chi _s }^d= {\chi _o}^{d \pm } |t| ^ {- \bar{\gamma} } \quad .
\label{chi_dis}
\end{equation}
The staggered magnetization, the order parameter for
antiferromagnets, is given by
\begin{equation}
M_s = M_o |t| ^ \beta \quad ,
\label{Ms}
\end{equation}
which is only nonzero for $t<0$.  In these expressions
$+$ and $-$ are for $t>0$ and $<0$, respectively.  The exponents
and the ratios for amplitudes above and below $T_c$ are universal
parameters common to all systems sharing the same symmetries.
The asymptotic critical exponents satisfy scaling relations such as
\begin{equation}
\gamma + \alpha + 2 \beta = 2 \quad .
\label{scaling}
\end{equation}
There are also hyperscaling
relations that involve the dimension $d$ such as
\begin{equation}
\alpha + \nu d = 2
\label{hyper}
\end{equation}
that hold for pure and REIM systems but are violated in the RFIM,
in which case Eq.\ \ref{hyper} is modified~\cite{v85,f86} by the
additional violation-of-hyperscaling exponent, $\theta$, with
\begin{equation}
\alpha + \nu (d- \theta ) = 2 \quad .
\label{violation}
\end{equation}

As the RFIM transition is approached by varying $H$ or $T$,
one observes a crossover from the zero-field universality
class to the RFIM one.  The crossover
behavior can be described by a crossover scaling
function.  For example, the free energy is given by
\begin{equation}
F=|t| ^ {2- \alpha } f(t h_r ^ {-2/ \phi } ) \quad ,
\label{scalefunct}
\end{equation}
where $ \phi $ is the crossover exponent, $\alpha$ is the
zero-field exponent and $h_r$ is the random-field strength.
A consequence of crossover between
different universality class behaviors is that
measurements may not yield asymptotic universal
parameters unless the data are sufficiently close
to $T_c$.  Rather, one obtains effective exponents.
The scaling relations (not the hyperscaling ones)
between exponents are still approximately
satisfied by the effective exponents~\cite{aa80}.
Another consequence of the crossover function is
a depression of the phase transition temperature given by
\begin{equation}
T_c(H)=T_N-AH^{2/ \phi}-bH^2 \quad ,
\label{phase-boundary}
\end{equation}
where $b$ represents a small mean-field shift
also present in the pure system.
The H-T phase boundary curvature is determined by $\phi$.
Fishman and Aharony~\cite{fa79} showed that
for the crossover from {\em pure} to random-field $d=3$
behavior, $\phi=\gamma$, with $\gamma = 1.25$ obtained
from theory and experiment~\cite{by87}.
Although some of the early experiments~\cite{sof84,wmd83,wmd82}
were incorrectly interpreted as showing this,
it was also argued~\cite{bkjc83}
that $\phi$ is much larger.  The latter result now appears to be universal,
with $\phi=1.42+0.03$ obtained for $Fe_xZn_{1-x}F_2$~\cite{fkj91},
$\phi=1.43 \pm 0.03$ for $Mn_xZn_{1-x}F_2$~\cite{rkj88} and
$\phi=1.41 \pm 0.05$ for $Fe_xMg_{1-x}Cl_2$~\cite{lk84}.
Aharony ~\cite{a86} predicted that for a {\em random-exchange} to random-field
crossover, $\phi$ is several percent larger than $\gamma$.
Neutron scattering measurements~\cite{bkj86} in
$Fe_{0.46}Zn_{0.54}F_2$ yielding $\gamma=1.31 \pm 0.03$
confirm this.  This was similarly verified~\cite{mcybub?}
in $Mn_xZn_{1-x}F_2$ with the result $\gamma = 1.36 \pm 0.08$.
The early interpretations~\cite{sof84,wmd83,wmd82} that $\phi=1.25$ were
influenced by the concentration gradients in the samples~\cite{kfjb88}
and the resulting misidentification of $T_N$.

The scaling function has other consequences for
random-field antiferromagnets.  Fishman and Aharony~\cite{fa79}
obtained
\begin{equation}
kT\chi =A_1+A_2|t|^{1-\alpha}-A_3|t|^{2\beta}
\label{suscept-pure}
\end{equation}
for the static $H=0$ uniform
susceptibility for a system dominated by the {\em pure} critical
exponents $\beta$, $\alpha$ and $\gamma$ at $H=0$.
The experimental systems, however, are dominated by {\em random-exchange}
critical behavior at $H=0$ and the correct relationship is therefore~\cite{a86}
\begin{equation}
kT\chi =A_1+A_2|t|^{1-\alpha}-A_3|t|^{2 - \alpha - \phi} \quad ,
\label{suscept-reim}
\end{equation}
using the REIM exponent $\alpha$ and 
the REIM to RFIM crossover exponent $\phi$.
For $H>0$, Kleemann, et al.~\cite{kkj86} showed,
by considering leading singularities in
derivatives of the free energy for $H \ne 0$,
that the field dependence of the amplitude of the
peaks in the specific heat, 
$(\partial M/ \partial T)_H$ and $( \partial M/ \partial H)_T$ is governed
by the exponents $y=(2/\phi)(\tilde{\alpha} - \alpha)$,
$y=(2/\phi)(1+\tilde{\alpha} - \alpha - \phi /2)$, and
$y=(2/\phi)(2+\tilde{\alpha} - \alpha -\phi )$, respectively,
where $\tilde{\alpha}$ is the RFIM specific heat exponent.
The exponents can be determined from specific heat~\cite{sbf96}
and Faraday rotation~\cite{kkj86} experiments on $Fe_xZn_{1-x}F_2$
to be $y \approx 0.1$, $y=0.60 \pm 0.10$,
and $y \approx 0.97 $, respectively.  For $Fe_{0.7}Mg_{0.3}Cl_2$,
$y=0.56 \pm 0.05$~\cite{lkf88b} for $( \partial M/ \partial H)_T$.
Since the exponents
$\phi = 1.42 \pm 0.03$ and $\alpha = -0.09 \pm 0.03$ are known
fairly accurately, we can invoke these results as a strong indication
from scaling that $\tilde{\alpha} \approx 0$ in good agreement with
direct specific heat results discussed below.  There is ample evidence
that scaling works well in all of these systems, despite recent
arguments~\cite{bfhhrt95} to the contrary.  Hence, at this time, experimental
interpretations should be constrained to agree with scaling theory.

\section{The $d=3$ RFIM Transition}

The first evidence that a phase transition occurs in the $d=3$ Ising
model came from the critical behavior of the specific heat measured using
optical linear birefringence~\cite{bkjc83} $(\Delta n)$ techniques,
which minimize the effects of concentration gradients since the laser
beam is directed perpendicularly to the concentration gradient.  The
proportionality~\cite{fl68,fg84} between $\frac {d(\Delta n)}{dT}$
and the magnetic contribution $(C_m)$ to the pulsed specific heat $(C_p)$
data has been shown explicitly~\cite{sbf96,db89}
for $Fe_{0.46}Zn_{0.54}F_2$ and $Fe_{0.93}Zn_{0.07}F_2$.
In anisotropic systems the Zeeman contribution
to the birefringence is negligible at
reasonable fields~\cite{fg84}.
In addition, Faraday rotation~\cite{kkj86} and
susceptibility~\cite{bkk95} measurements yield the
specific heat critical behavior.
Recent claims to the contrary~\cite{bfhhrt95,w96}
have no theoretical basis~\cite{fl} and depend on
analyses of experimental data which have been questioned~\cite{bkm96}.

The specific heat critical behavior in the intermediate
range, $0.4< x < 0.8$, has been measured in
$Fe_xZn_{1-x}F_2$~\cite{sbf96,pkb88} and $Fe_xMg_{1-x}Cl_2$~\cite{lkf88b}
with enough precision to determine that
ZFC data are well described by a symmetric
logarithmic divergence over a reasonable
range in $t$.  At very
small $|t|$, rounding is observed and is attributed to
the tremendous critical slowing down of the RFIM in dilute antiferromagnets,
as will be discussed in the following section.  FC always
yields behavior that is much more rounded because finite-size metastable
clusters~\cite{bkk95,bk95} are frozen in just above $T_c(H)$.
The region over which the logarithmic ZFC behavior
and the dynamic rounding are observed
increases with $H$ as expected
from crossover scaling (Eq.\ \ref{scalefunct}) and dynamic scaling
as discussed in the next section.

\begin{figure}[t]
\centerline{\hbox{
\psfig{figure=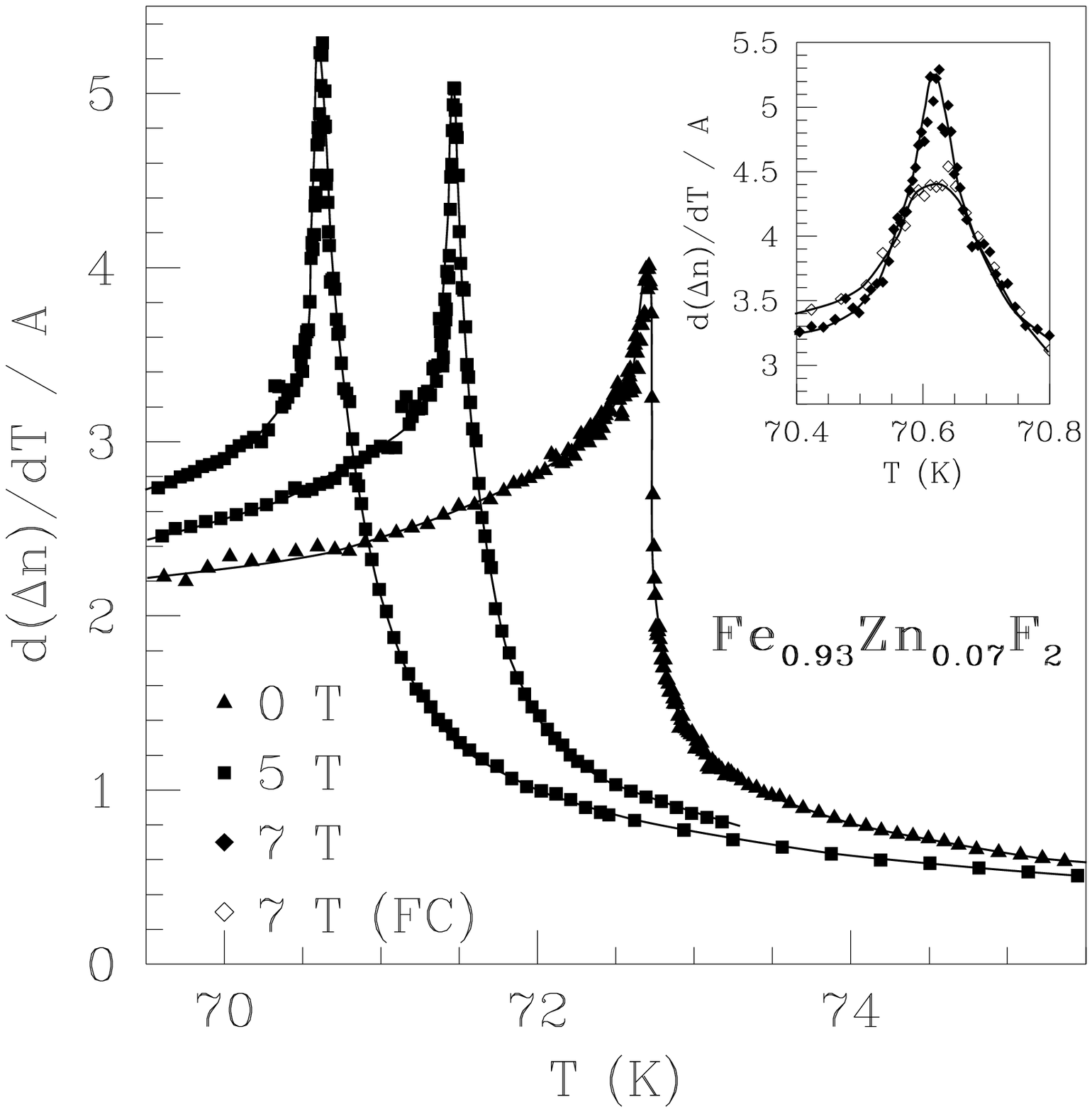,height=3.in}
\psfig{figure=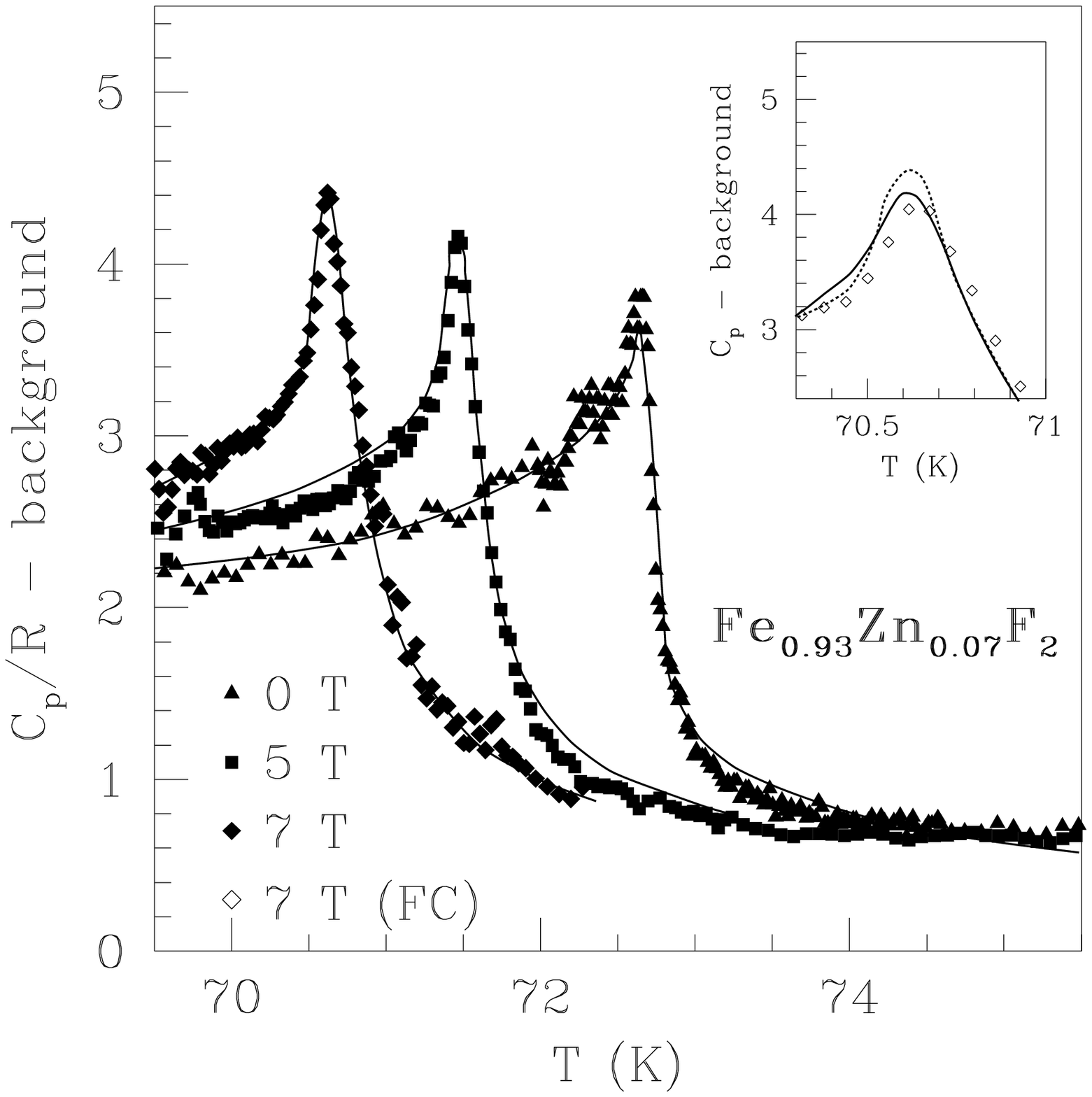,height=3.in}
}}
\caption{$ \frac {d(\Delta n)}{AdT}$ vs.\ $T$,
where $A=9.17 \times 10^{-6}$~K$^{-1}$ is the same
proportionality constant found for pure
$FeF_2$, and $C_m/R$ vs.\ $T$
for $Fe_{0.93}Zn_{0.07}F_2$.  The specific heat has the phonon
contribution subtracted as discussed in the text.  ZFC data are
shown in the main figures.  The insets show the $H=7$~T FC data as well as the
ZFC data for the $\frac {d (\Delta n)}{dT}$ case.  The curves in the
left figure are the same as the curves in the figure on the right except
that they are rounded by the larger, measured concentration
gradient.  For the
specific heat inset, the ZFC data are not shown, for clarity,
but the dotted line is the same as the solid ZFC line in the main figure.
Just as in experiments at lower concentrations, the birefringence
and pulsed heat techniques yield precisely the same behavior, including
FC-ZFC hysteresis very close to $T_c(H)$.
The critical behavior for $H>0$ is closely approximated by a symmetric,
logarithmic divergence.
}
\end{figure}

Figure 1 shows recent $\frac {d (\Delta n)}{dT}$
and $C_m$ data for a high magnetic
concentration sample~\cite{sbf96} $Fe_{0.93}Zn_{0.07}F_2$.  We have
subtracted the approximate large phonon contribution to $C_p$,
leaving only $C_m$, so that the correspondence of the two techniques
is apparent.  We accomplished this by assuming the same
proportionality~\cite{db89} between $C_m$ and $\frac {d(\Delta n)}{dT}$
found for both $FeF_2$ and $Fe_{0.46}Zn_{0.54}F_2$.  The excess specific heat
contribution found for the $H=0$
was then subtracted from all the $C_p$ data sets.
The solid curves in the $C_m$ figure are adapted from those
in the $\frac {d (\Delta n)}{dT}$ one by first numerically rounding
the $\frac {d (\Delta n)}{dT}$ data by the known gradient, drawing
smooth curves through them,
and then transferring the curves to the $C_m$ figure with
no other adjustments.
The $C_m$ peaks are not as sharp since the entire sample
was used, increasing the effects of concentration gradients relative
to the $\frac {d(\Delta n)}{dT}$ data which are sensitive only to the
gradient along the laser beam.  Clearly, the known gradient
accounts well for the difference in the $C_m$ and $\frac {d(\Delta n)}{dT}$
data.  The insets for both sets of data show the FC behavior at $H=7$~T.
The FC data are more rounded than the ZFC.
The FC $C_m$ data are shown in the inset on the right
with the $\frac {d(\Delta n)}{dT}$ FC curve, rounded
by the concentration gradient in the same manner as the
ZFC ones, shown as a solid curve.  The FC curve
corresponds well with the $C_m$ data.  The dotted curve corresponds to the
ZFC and is the same as the solid one in the main
figure.  For clarity, we do not
show the ZFC data in the inset.  No other adjustments have been made.
Clearly, the hysteresis is much more difficult to discern in
the specific heat data~\cite{bkm96}, but this is consistent with the larger
concentration gradient.  The hysteresis can only be observed in
samples with extremely small gradients.  This is certainly one reason why some
experiments on samples with appreciable concentration gradients fail
to exhibit hysteresis.  In all respects we see that
the $\frac {d(\Delta n)}{dT}$ and $C_m$ data yield the same critical behavior
just as was found previously~\cite{db89} for $Fe_{0.46}Zn_{0.54}F_2$.
As discussed below, $Fe_{0.93}Zn_{0.07}F_2$ yields neutron scattering
line shapes that are fundamentally different from those obtained at lower
magnetic concentrations in that they do not show hysteresis well below
$T_c(H)$.  Yet, the specific heat appears remarkably similar to that of
lower magnetic concentrations.  The symmetric, logarithmic ($\alpha =0$)
behavior for ZFC contrasts with the $H=0$ data that are consistent with
the expected asymmetric random-exchange~\cite{h74} cusp.

For all concentrations there is a temperature, $T_{eq}(H)$, below which
hysteresis between the FC and ZFC procedures plays a role in the specific heat
as well as most other experiments.  We emphasize that $C_m$ and
$\frac {d (\Delta n )}{dT}$ data show precisely the same hysteresis,
contradicting recent claims~\cite{bfhhrt95}.  Using a capacitance
technique~\cite{kjbr85} on $Fe_{0.46}Zn_{0.54}F_2$ and $Fe_{0.72}Zn_{0.28}F_2$,
the equilibrium boundary $T_{eq}(H)$ has been shown to lie
just above $T_c(H)$, scaling precisely in the same manner with
$\phi \approx 1.42$.
The nature of this boundary is still not entirely clear, though it is
sharp enough to be measured precisely.  It could be
related to the extreme critical dynamics discussed in the next section or
it could be related to a RFIM spin-glass-like behavior~\cite{mm94,yb84}
between $T_{eq}(H)$ and $T_c(H)$.  Certainly one must be
careful about the data extremely close to $T_c(H)$ since
the system could be out of equilibrium.

Hysteresis in the specific heat is not well observed in
ac techniques~\cite{i86} used on the less anisotropic
system $Mn_xZn_{1-x}F_2$ at $6.6$ Hz.
The extremely rapid heating and
cooling method ($10$~K per minute)
of measuring specific heat in $Fe_xMg_{1-x}Cl_2$
also shows very little hysteresis~\cite{wmd82},
although early neutron scattering~\cite{wc83} and
measurements in this system
clearly exhibit hysteresis.  Perhaps the time dependent
techniques obscure the difference between
FC and ZFC, though this is not yet clear.
Recently, it was claimed that in $Fe_{0.5}Zn_{0.5}F_2$ no
hysteresis is observed in the specific heat~\cite{bfhhrt95}.
Although there is no published description of the procedures used, 
some conjectures can be made as to why the hysteresis was
missed.  Perhaps the sample concentration gradient induced rounding of
$0.3$ K~\cite{bfhhrt95} obscures the transition at the low field~\cite{bkm96}.
The phase boundary might have been exceeded at the 
high field~\cite{bkm96,llm96}.  Finally, if the measurements were not
sufficiently adiabatic, the hysteresis may be obscured as they appear to be in
other time dependent measurements~\cite{i86,wmd82}.
The answer is simply unclear at this time and the failure
to observe hysteresis could be a combination
of effects.

Whereas all the high resolution specific heat measurements done
to date indicate a symmetric, logarithmic
divergence with no evidence for any accompanying
background discontinuity~\cite{kk85},
Monte-Carlo simulations~\cite{ry93,r95} indicate
a cusp, with a large, negative exponent.
This discrepancy between the exponents from simulation and
experiment is, as yet, unresolved and is certainly a major challenge
to be addressed.

In contrast to the birefringence measurements that
first showed evidence~\cite{bkj82,bkjc83} of a d=3 transition, early neutron
scattering measurements obtained with the FC process
were interpreted as indicating a destroyed transition~\cite{ycsbgi82}.
Upon FC, no Bragg scattering is observed for concentrations $x < 0.8$.
Instead, a finite-width shape approximated
by a squared-Lorentzian, as discussed below, appears.
We now know that long-range antiferromagnetic order is difficult
to establish upon FC at low concentraton, but that a phase transition
is nevertheless the basis of the underlying physics.  Long-range
order does occur for $T<T_c(H)$ when the field is applied after ZFC and
FC domains have been shown to be metastable~\cite{bkj85c}.

Quite different phenomena are observed at high 
magnetic concentration.
Recent scattering measurements~\cite{sbf96}
using a crystal of $Fe_{0.93}Zn_{0.07}F_2$
indicate a ZFC transition that is as sharp as allowed
by the concentration gradient $\delta x = 0.002$.
More importantly, there is no evidence of nonequilibrium
hysteresis except extremely close to $T_c(H)$, as in the specific
heat~\cite{sbf96}.
What is most remarkable is that the $Fe_{0.93}Zn_{0.07}F_2$
neutron scattering line shapes show little hysteresis at low $T$.
For $x<0.8$, such hysteresis has always been observed and has been
a major obstacle to interpreting the critical scattering below $T_c(H)$.
An important distinction can be made between hysteresis seen
in line shapes at lower concentration well below $T_c(H)$ which most likely
originates in the large number of
vacancies, and the rounding near $T_c(H)$ that appears even at high magnetic
concentration.  The latter may well have to do with RFIM critical
dynamics as is clearly the case with the specific heat behavior~\cite{pkb88}.

The interpretation of the scattering results in RFIM studies
is severely hampered by the lack
of adequately characterized line shapes provided by theory.
As previously reviewed in more detail~\cite{by92,b88}, mean-field theory
yields an elastic scattering cross section of the form
\begin{equation}
S(q)= \chi (q) + {M_s}^2 \delta (q) = \frac{A} {q ^ 2 + \kappa ^ 2 } + {M_s}^2 \delta (q) ,
\label{lor}
\end{equation}
for a pure system and, with an additional squared-Lorentzian term,
\begin{equation}
S(q)=  \chi (q) +  \chi ^d (q) = \frac{A} {q ^ 2 + \kappa ^ 2 } 
+  \frac{B} {(q ^ 2 + \kappa ^ 2 )^2 } + {M_s}^2 \delta (q) ,
\label{lor_lor2}
\end{equation}
for a random-field system~\cite{l84}.
These expressions can be only approximate for $d=2$ or $d=3$
in pure or random systems, as one can see from the
required asymptotic behaviors
$\chi (0) \sim \kappa ^{2- \eta } \sim |t|^{- \gamma }$ and
$\chi ^d (0) - {M_s}^2 \sim \kappa ^{4- \bar{\eta}} \sim |t|^{- \bar{\gamma}}$.
The correspondence between the
measured line shapes and the mean-field line shapes in pure
systems is fairly good for $d=3$ since $\eta \approx 0.04$ is small,
though evidence for deviations from mean-field behavior have
been observed~\cite{by87}.  Although
Pelcovits and Aharony~\cite{pa85} predict
significant deviations from the Lorentzian line shape for $T<T_c(H)$
in the $d=3$ REIM, where $\eta$ is also small, no
definitive evidence for this has yet been observed in experiments.
For $d=2$, the discrepancy between the line shapes of the pure system
and mean-field theory is more evident~\cite{fb67,by87} since $\eta = 1/4$.
For random-field systems, $\eta \approx 1/2$ is large~\cite{by92} and the
mean-field terms in Eq.\ \ref{lor_lor2} are expected to be far from
accurate.  The observed line shape in the random-field
systems is in many cases inconsistent with the Lorentzian
in Eq.\ \ref{lor}, as was first shown by Yoshizawa, et al.~\cite{ycsbgi82}.
However, the story is not as simple as
adopting Eq.\ \ref{lor_lor2} since this expression is often inconsistent
with the data~\cite{bkjn87}, particularly below $T_c(H)$.
Nevertheless, Eq.\ \ref{lor_lor2} is a start.

\begin{figure}[t]
\centerline{\hbox{
\psfig{figure=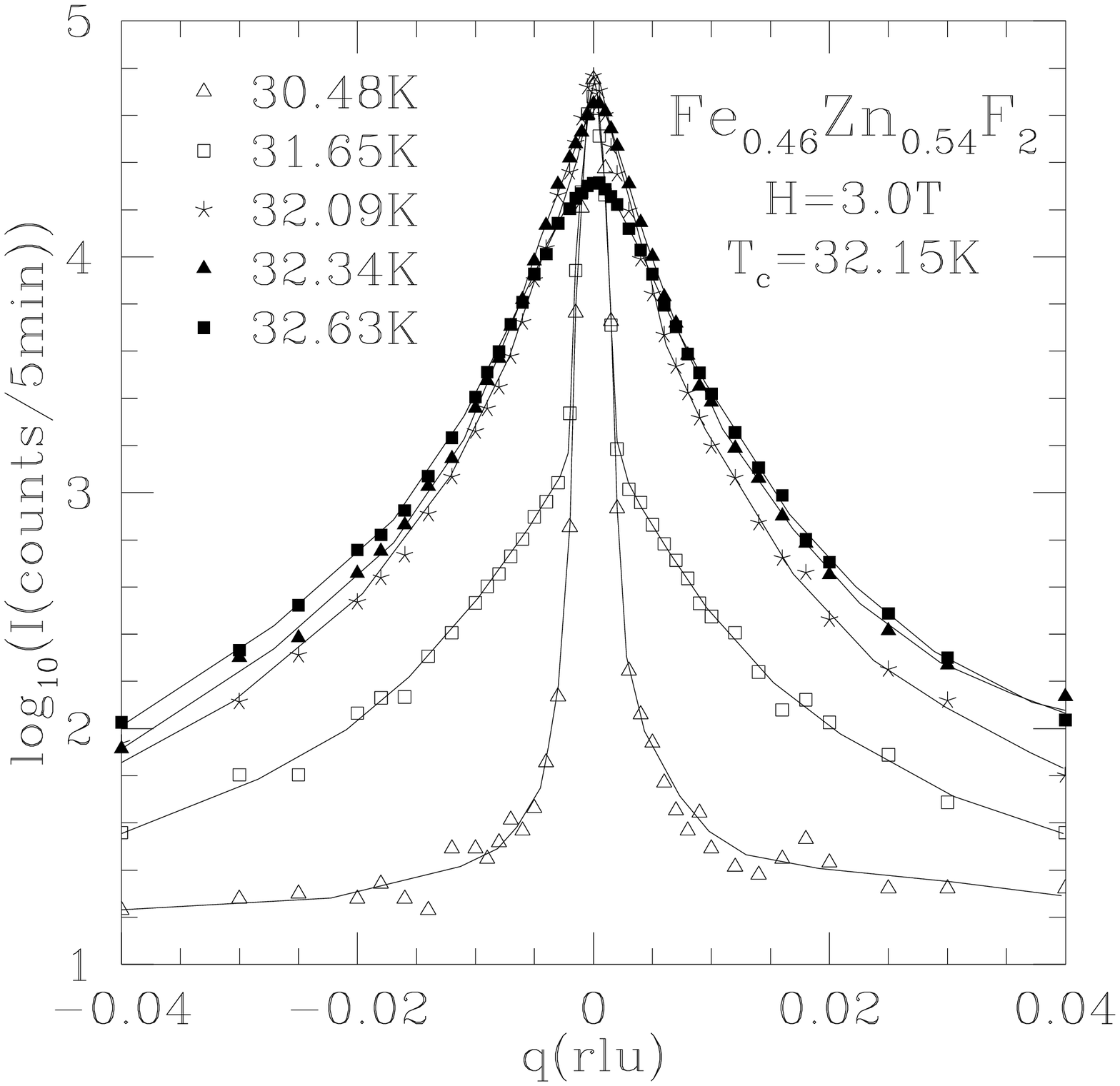,height=2.9in}
\psfig{figure=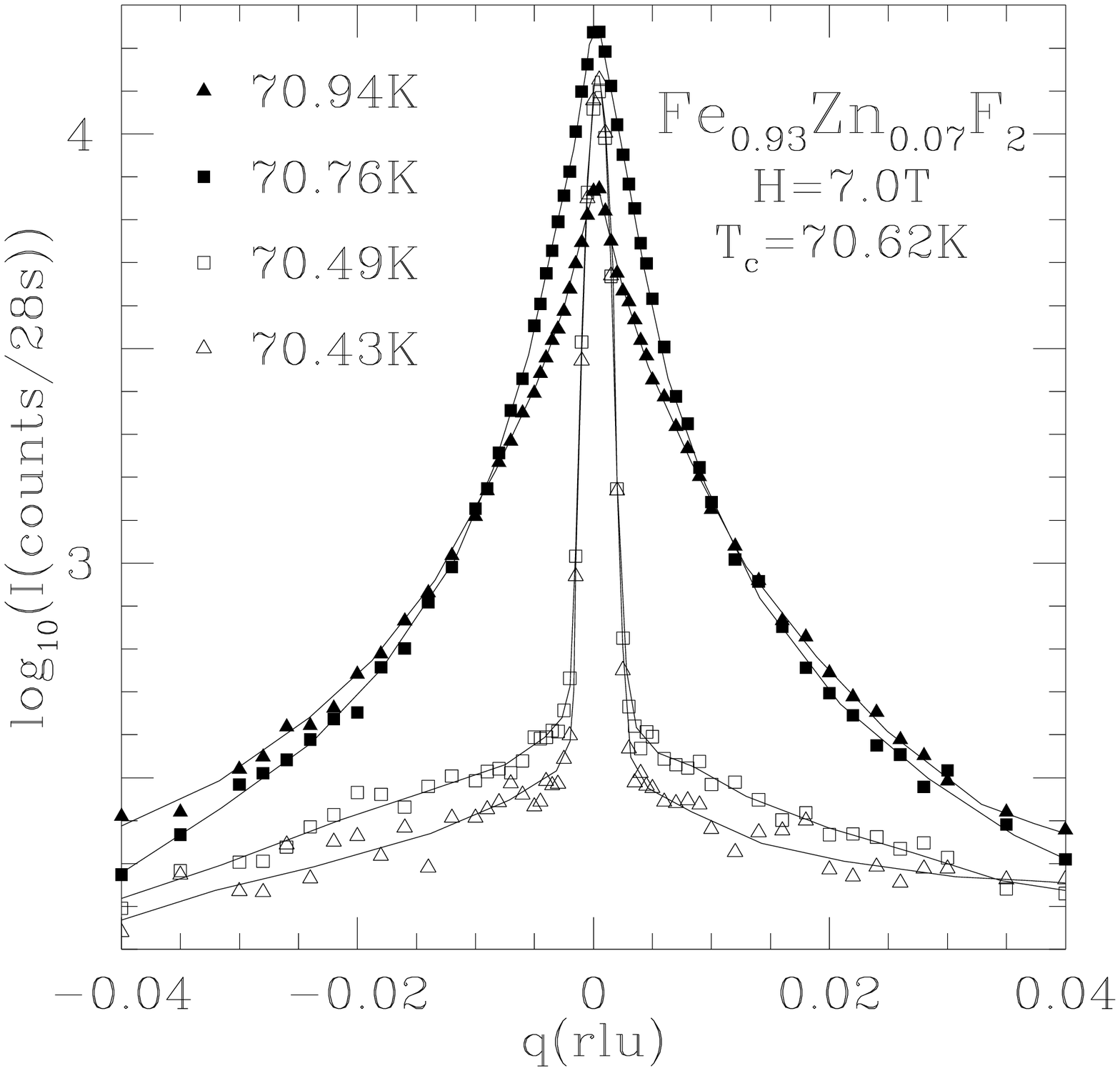,height=2.9in}
}}
\caption{Neutron scattering intensity, $I(q)$ vs.\ $q$ for
$Fe_{0.46}Zn_{0.54}F_2$ and $Fe_{0.93}Zn_{0.07}F_2$ after ZFC.
Above $T_c(H)$, determined from the peak in the critical
scattering, both samples exhibit Lorentzian plus squared-Lorentzian
line shapes.  Below $T_c(H)$, the $Fe_{0.46}Zn_{0.54}F_2$ crystal
shows a resolution limited lineshape which conforms to neither
a Lorentzian nor a squared-Lorentzian line shape in addition to
the Bragg scattering peak.  The Bragg scattering decreases
dramatically at this concentration as $T_c(H)$ is approached and
shows a large hysteresis upon temperature cycling below $T_c(H)$.
This behavior is consistent with the system breaking into large,
intertwined, weakly interacting domains, a result
of the very large number of vacancies at this concentration.
In contrast, the $Fe_{0.93}Zn_{0.07}F_2$ crystal exhibits
Lorentzian line shapes below $T_c(H)$ with no sign of domain formation.
No hysteresis for $|q|>0$ is observed well below $T_c(H)$, indicating
equilibrium behavior.  The Bragg intensity remains large extremely close
to $T_c(H)$, indicating
that $\beta$ is very small, in agreement with simulations.
}
\end{figure}

The first neutron scattering observations of the $d=3$ RFIM
phase transition~\cite{bkj85b}, made using $Fe_{0.6}Zn_{0.4}F_2$,
showed that above the transition
the ZFC line shapes are inconsistent with the single Lorentzian
term but can be fit adequately by the sum
of Lorentzian and squared-Lorentzian terms as in Eq.\ \ref{lor_lor2}.
Non-Lorentzian line shapes had been observed previously~\cite{ycsbgi82}
for the metastable domain state well below $T_c(H)$ after FC.
From the fits to the critical behavior above $T_{eq}(H)$,
the estimations $\nu = 1.00 \pm 0.03$,
$\gamma = 1.75 \pm 0.20$ and $\bar{\gamma} = 3.5 \pm 0.3$ are
obtained~\cite{bkj85b}.  An attempt at a better characterization,
including for $T<T_c(H)$, was made~\cite{bkjn87}
with the very homogeneous crystal
$Fe_{0.46}Zn_{0.54}F_2$.  Several scans are shown in Fig.\ 2.
Although the ZFC scattering above $T_c(H)$ is
indeed fairly well fit by Eq.\ \ref{lor_lor2}, the scattering line shapes
below $T_c(H)$ are certainly not.  Below $T_c(H)$ the measured scattering
profiles are much too
narrow, being essentially resolution limited for all $T<T_c(H)$
instead of having a width that varies as $\kappa (T)$.  Furthermore,
the intensity of the Bragg component is surprisingly small
near $T_c(H)$ and qualitatively
it appears as if the Bragg peak transforms into the non-Lorentzian,
resolution-limited
scattering contribution as $T_c(H)$ is approached from below.  Finally,
a peak in the $q=0$ scattering intensity  is observed~\cite{bkjn87} at $T_c(H)$
upon ZFC and the height of the peak grows approximately logarithmically
with time, a result of the extremely slow dynamics.
Such a peak is difficult to observe normally but is evident in
this case because the Bragg scattering, which usually swamps
the critical fluctuation peak, is abnormally small just below $T_c(H)$.

Although it is clear that the scattering is peculiar and interesting,
extinction effects~\cite{cyshb84} give pause to direct interpretations
of the Bragg scattering intensities in the $Fe_{0.46}Zn_{0.54}F_2$
experiments.  High quality crystals scatter neutrons that are precisely
aligned for the Bragg scattering condition in the first
ten microns or so of material.  As the scattering cross section
diminishes upon approaching $T_c(H)$, the scattering
simply occurs over a larger volume.  Hence, the scattered intensity
is saturated and does not exhibit the power law behavior
in Eq.\ \ref{Ms}.  This difficult problem has been overcome by
examining~\cite{bwshnlrl96} an epitaxial~\cite{lkjs89}
$Fe_{0.52}Zn_{0.48}F_2$ film
of thickness $3.4$ $\mu$m, grown on a $ZnF_2$ substrate.
The film is thin enough to avoid extinction effects
but thick enough ($ \approx 10^4 $ lattice spacings) for $d=3$
critical behavior.

The neutron scattering results for the $H=0$
Bragg intensity~\cite{bwshnlrl96}
of the film are consistent with REIM behavior.  Hence,
the film is high enough~\cite{lkjs89} in quality to reliably reflect
the $d=3$ critical behavior.  The scattering
intensity for $q>0$, coming solely from a Lorentzian contribution,
is too weak to be observed in the film.
For $H>0$ the scattering results are highly unusual.
The ZFC Bragg intensity vs.\ $T$ has the opposite curvature
to that observed for $H=0$, so the Bragg scattering intensity is very
small quite far below $T_c(H)$.  The loss of the ZFC Bragg intensity
is irreversible below $T_c(H)$.  This behavior has been interpreted
as the system breaking into two intertwined domains with equal numbers
of spins in a similar pattern to that observed in FC
simulations~\cite{nu91} at low $T$.
The formation of domains is observed to be irreversible below $T_c(H)$,
a result that is consistent with the irreversibility observed in
magnetization and optical studies\cite{lkf88a,pkb88,lk84}.
Inside the domains the spins are well ordered.  The domain
walls at this concentration ($x \approx 0.5$) are able to
pass predominantly through the numerous vacancies, costing
the system very little energy.  It is clear that the
Imry-Ma domain wall energy arguments~\cite{im75} fail here since the energy
needed to create such a domain wall is insignificant compared
to the Zeemann energy decrease.  Furthermore,
the domains are only weakly interacting and each contributes
to the phase transition at $T_c(H)$.  Since the domains form well below
$T_c(H)$, neutron scattering measurements are unable at this
concentration to determine the critical behavior of the
order parameter.  Another piece of evidence indicating that the
hysteresis for $T_c(H)$ comes from domain formation is found in
the experimental
results of x-ray scattering studies~\cite{hfbt93} at the surface
of $Mn_{0.75}Zn_{0.25}F_2$.  In the presence of surface defects,
no hysteresis is observed, most likely a result of the defects
preventing the formation of the two intertwining domains.  When an
identical sample was polished, removing the majority of defects,
the hysteresis reappeared.

In both the film ($Fe_{0.52}Zn_{0.48}F_2$) and bulk
($Fe_{0.46}Zn_{0.54}F_2$) studies, we find a large
resolution-limited scattering line shape below the
transition that is not well fit by either a Lorentzian
or squared-Lorentzian term.  It is most likely that 
this non-Lorentzian scattering profile is a signature of domain
structure that forms below $T_c(H)$ even upon heating after ZFC.
With this structure present it is very difficult to determine
the critical behavior of the RFIM below the transition.
Local probes like NMR~\cite{mskjiyh86}, M\"{o}ssbauer or $\mu$SR
in principle could yield the order parameter critical
behavior, but prove to be complicated because of the spatial
variations within the system.
This motivated an investigation at a much higher concentration,
where the vacancy concentration is
small enough that domain walls cannot easily avoid a
large energy cost of formation.  Since hysteresis at
low temperatures is seen in the work~\cite{bcsy85}
on $Mn_{0.75}Zn_{0.25}F_2$, it is clear that one must go to even
higher concentrations.  Preliminary measurements~\cite{sbf96} using the
$Fe_{0.93}Zn_{0.07}F_2$ crystal seem to confirm the idea; the hysteresis
in the scattering profile at low temperatures is eliminated.

The abrupt change in line shape of $Fe_{0.93}Zn_{0.07}F_2$ at $T_c(H)$
is striking.  Figure 2 shows scans
taken just above and just below $T_c(H)$ at $H=7$~T.
Just 0.13K below $T_c(H)$ the line shape is incompatible
with any significant squared-Lorentzian term.  A Lorentzian
term fits fairly well.  The absence of the
non-Lorentzian component is most likely a signature of
the stability of the
long-range order right up to $T_c(H)$.  Above $T_c(H)$, on the other
hand, the line shapes are much more compatible with a fit to a Lorentzian
plus squared-Lorentzian as in Eq. \ref{lor_lor2}.
The abrupt disappearance of the Bragg peak at $T_c(H)$, indicating
a very small value for $\beta$, contrasts greatly the behavior observed
at lower concentrations in $Fe_xZn_{1-x}F_2$.
A small value of $\beta $ is consistent with theory and
simulation results~\cite{ry93,r95,yn85}.  The only previous
experimental measurement~\cite{rkjr88} of $\beta$
is from dilation experiments on the lower concentration
sample $Fe_{0.46}Zn_{0.54}F_2$ which indicates $\beta \leq 1/8$.
This suggests that the small
exponent value holds for lower concentrations even though the neutron
scattering Bragg intensity cannot show it.
The small value of $\beta$ is perhaps suggestive of a first-order transition,
but no latent heat is observed in the specific heat
in the experiments or simulations~\cite{sbf96,db89,yn85,lmp95}.

At the lower magnetic concentrations, severe hysteresis
is observed in the line shapes below $T_c(H)$.  In the case of
$Fe_{0.93}Zn_{0.07}F_2$, however, the line shapes for $q>0$ do
not exhibit hysteresis except for the region near $T_c(H)$
where critical dynamics dominate.  The Bragg intensity
does show some hysteresis, being somewhat larger upon FC, but this
is an extinction effect~\cite{cyshb84} reflecting the fact that
long-range order on length scales well beyond the instrumental
resolution is not established upon FC, most likely a result of
RFIM dynamics very close to $T_c(H)$.  The temperature dependence
of the Bragg intensity is essentially  the same for the Bragg intensity upon
ZFC and FC well below $T_c(H)$.

\begin{figure}[t]
\centerline{\hbox{
\psfig{figure=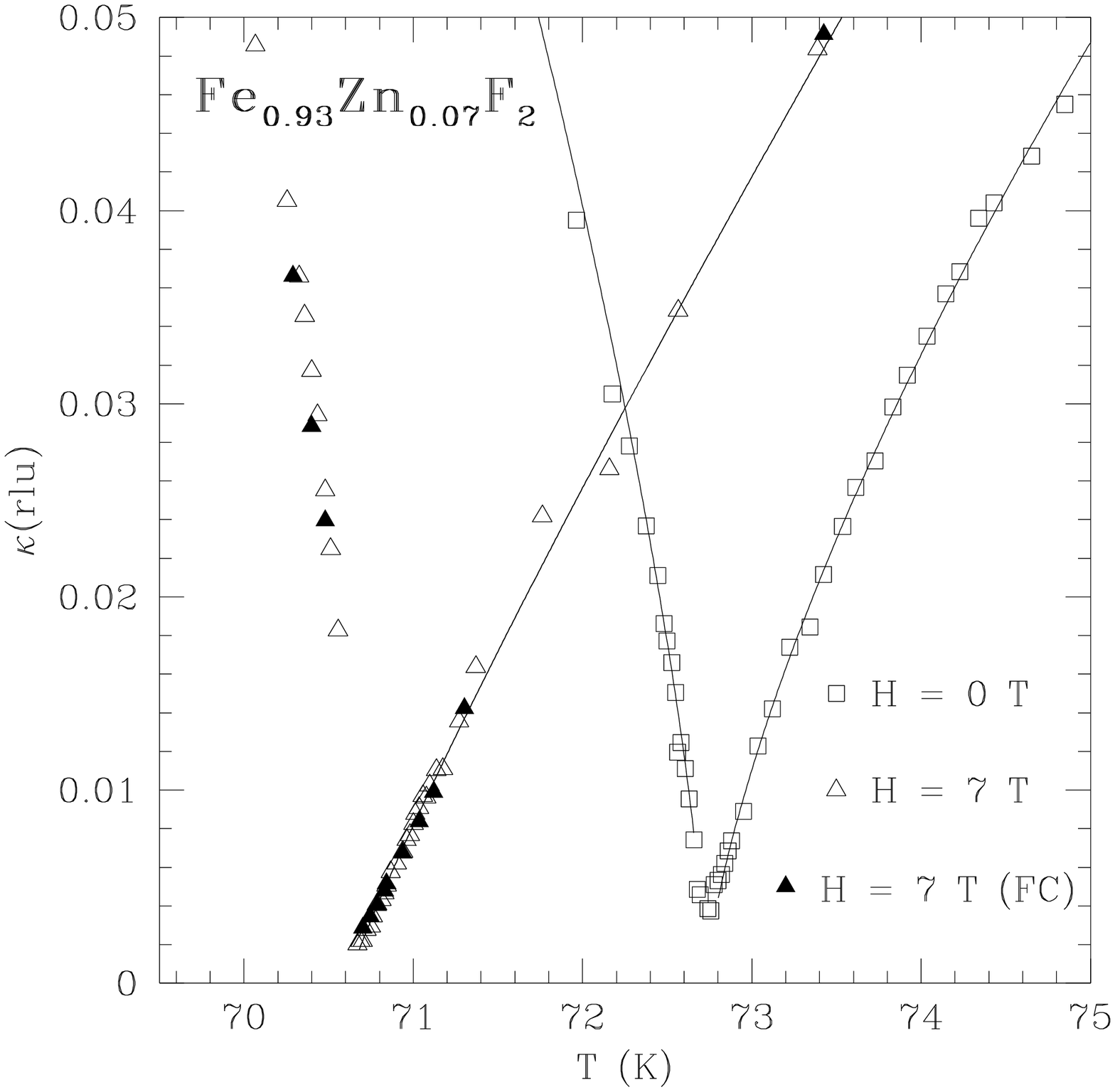,height=3.in}
\psfig{figure=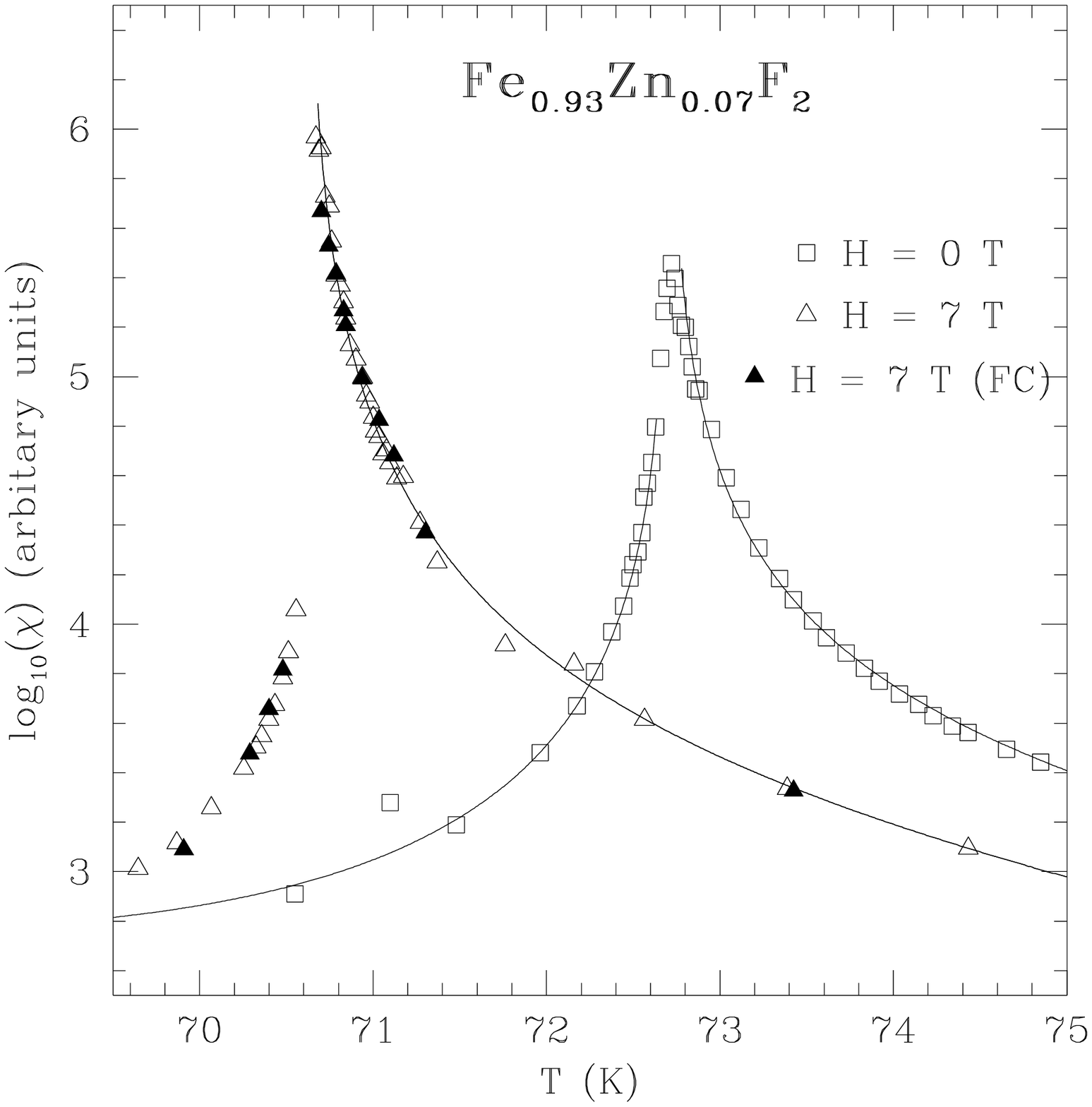,height=3.in}
}}
\caption{$\kappa $ vs.\ $T$ and $\chi $ vs.\ $T$ for
$Fe_{0.93}Zn_{0.07}F_2$ for $H=0$ and 7~T, obtained
from preliminary analysis of the neutron scattering
line profiles for $|q|>0$.  For $H=7$~T and $T>T_c(H)$,
the $|q|>0$ scattering is fit to a Lorentzian plus squared-Lorentzian
lineshape.  A Lorentzian was used in all other cases.
For $H=7$~T, the
open triangles are for ZFC and the filled ones are for FC.  The lack
of hysteresis indicates equilibrium behavior.
The solid curves are fits to the data.  However, for $H=7$~T and $T<T_c(H)$,
no power law describes the data well, so no curve is shown for this case.
}
\end{figure}

Evidently, if we can extract the critical behavior in
the $Fe_{0.93}Zn_{0.07}F_2$ sample, it should represent
the equilibrium behavior since it is history independent.  Unfortunately,
the lack of a theoretical scattering line shape that goes beyond the
misleading mean-field theory of Eq.\ \ref{lor_lor2} has severely limited
the extraction of critical parameters below $T_c(H)$.  Experimental work in
this area is ongoing with progress anticipated, but theoretical
work is also much needed in the near future.  Above the transition
the fits to Eq.\ \ref{lor_lor2} seem to work fairly well and
one can extract the exponents,
albeit with trepidation regarding exact results.
The results for $\kappa$ and $\chi$ vs. $T$ are shown
in Fig.\ 3 along with fits represented by the solid curves.
Fits were made for all of the $H=0$ data and for $T>T_c(H)$ with the
$H=7$~T data.  No suitable fit to a power law is obtained for
$T<T_c(H)$ and no curves are shown. 
Preliminary fits~\cite{sbf96} for $T>T_c(H)$ yield $\nu = 0.93 \pm 0.03$,
$\gamma = 1.71 \pm 0.06$ and $\bar{\gamma} = 3.0 \pm 0.1$
for $10^{-3}<t<10^{-2}$.
These values are in reasonable agreement with earlier experimental
results~\cite{bkj85b} at $x=0.6$ mentioned above
but are in disagreement with other estimations where the transition
appears distinctly rounded~\cite{cbsy} from concentration gradients.
(Larger values for $\nu$ have been obtained in other studies, but only because
$T_c(H)$ has been taken to be well below the minimum in $\kappa$ in
samples with relatively large gradient induced rounding.)
There is reasonably good agreement between the exponents obtained
from neutron scattering in $Fe_{0.93}Zn_{0.07}F_2$ and those obtained
from Monte Carlo simulations.   For example, Rieger~\cite{r95} obtains
$\nu =1.1 \pm 0.2$, $\gamma = 1.7 \pm 0.2$, $\bar{\gamma} = 3.3 \pm 0.6$,
and $\beta = 0.00 \pm 0.05$ for a Gaussian distribution of random fields.
The scattering results are also reasonably consistent with recent
high temperature expansion~\cite{gaahs93} results for $\gamma$ and
$\bar{\gamma}$.

Keeping in mind the uncertainty concerning the scattering line shape
appropriate for analyzing the $Fe_{0.93}Zn_{0.07}F_2$ data,
the preliminary scattering exponents above $T_c(H)$,
$\gamma = 1.71$ and 
and from the specific heat, $\alpha \approx 0$, satisfy the simple
scaling relation in Eq.\ \ref{scaling} if $\beta $ is small
as expected from theory.
In stark contrast, a typical result from Monte Carlo simulations is that
$\alpha$ is large and negative, for example
$\alpha = -0.5 \pm 0.2$~\cite{r95}.
Nevertheless, the specific heat exponent is the most consistent
experimental exponent.  Note that the measured
amplitude ratio $A^+/A^-$ is very close to unity which is consistent
with a logarithmic divergence.
Also, as demonstrated in section 3, amplitude scaling relations for dilute
antiferromagnets strongly indicate $\alpha \approx 0$.

We can use Eq.\ \ref{violation} and the measured
exponent $\nu=0.93$ for $T>T_c(H)$ to estimate the violation
of hyperscaling exponent $\theta = 0.85 $.
Using the relations $\gamma = \nu (2-\eta )$
and $\bar{\gamma}=\nu (4-\bar{\eta} )$ with the values from scattering
$\gamma = 1.71$ and $\bar {\gamma} = 3.0$, we can estimate
$\eta = 0.16 $ and $\bar{\eta} = 0.77$.
These values are smaller than theoretical estimates, but they are very
preliminary and further measurements and analysis
will certainly refine them in the near future.
The point to be made is that we are finally almost at the stage where
serious comparison with theory can be made, though we are greatly
hampered by not knowing the correct line shape.

Finally, we should briefly mention a very recent
suggestion by Birgeneau, et al.~\cite{bfhhrt95,hfbt93} that
the unusual curvature of the Bragg intensity versus $T$ is actually
a rounding of the phase transition at intermediate
concentrations - the ``trompe l'oeil'' phenomenological model,
as they have labelled it.  It was introduced in an attempt to describe the
scattering, magnetization and specific heat behavior
of the $d=3$ RFIM phase transition in the lower concentration antiferromagnets
$Fe_{0.5}Zn_{0.5}F_2$ and $Mn_{0.75}Zn_{0.25}F_2$.
The interpretation of the data in this model conflicts sharply with
the interpretations presented in this review~\cite{bkm96}, since it clearly
violates scaling for $H>0$, which was developed by Kleemann,
et al.~\cite{kkj86} and Fishman and Aharony~\cite{fa79} and is described in section 3.
The authors take this as evidence that the scaling theory is
incorrect.  The interpretation requires that the peak in
$(\partial M/ \partial T)_H$ coincide with the peak in
$( \partial {M_s}^2/ \partial T)_H$ and, to accomodate this,
the $( \partial {M_s}^2/ \partial T)_H$ data are adjusted within the
thermometry uncertainties.  The shifts of the data
weaken the motivation for the new model and the argument that scaling
fails.  The proposed model requires that the uniform
magnetization couple strongly to the antiferromagnetic long-range order
and this has not yet found theoretical motivation~\cite{fl}.  The model also
depends on the specific heat in $Fe_{0.5}Zn_{0.5}F_2$ showing no hysteresis,
but such hysteresis has been observed in pulsed heat experiments using
$Fe_{0.46}Zn_{0.54}F_2$~\cite{db89} and $Fe_{0.93}Zn_{0.07}F_2$~\cite{sbf96}
crystals with very small concentration gradients.

\section{Critical Dynamics of the $d=3$ RFIM Transition}

\begin{figure}[t]
{
\hspace{0.75in}
\psfig{figure=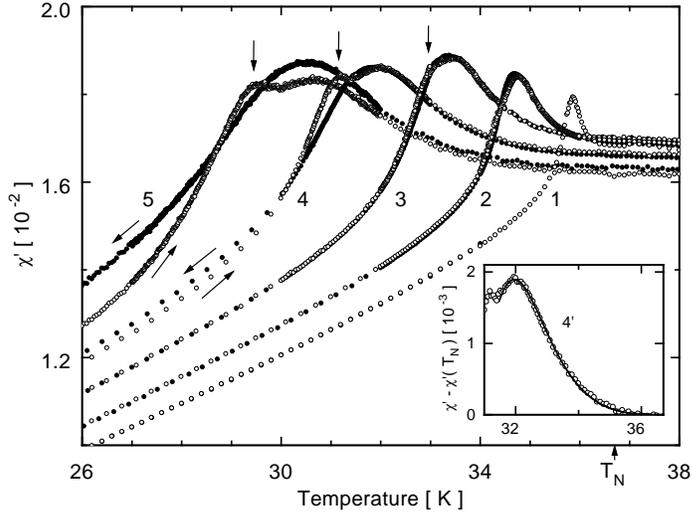,height=2.9in}
}
\caption{$\chi '$ vs. $T$ at $f=1$~Hz for $Fe_{0.47}Zn_{0.53}F_2$
for applied fields $H=0.8$, $1.6$, $2.4$, $3.2$ and $4$~MA/m.
Open circles are ZFC data and filled circles are FC.
The main contribution to the peak is
from Griffiths-like clusters which form above $T_c(H)$.
The smaller peak, which is resolved only at larger fields
and only upon FC,
is at the actual phase transition.  From the dependence
of the small ZFC peak height vs.\ the frequency, the dynamics
can be ascertained.  The behavior is consistent with
a power law with a very large dynamic exponent or with
activated dynamics.
The inset shows a fit to the broad peak at $H=3.2$~MA/m
using a phenomenological Griffiths-cluster model.
}
\end{figure}

The critical dynamics of the $d=3$ RFIM transition are
extraordinarily slow.  Many of the experiments
in $Fe_xZn_{1-x}F_2$ for $H>0$ that would normally be considered
static measurements have shown behavior with approximately logarithmic
time dependence.  These include
neutron critical scattering~\cite{bkjn87} and capacitance
techniques~\cite{kjbr85}.  Spin-echo neutron scattering techniques
show~\cite{bfjklm88} that very small fields suffice to freeze the
system over the entire critical region $|t|<0.1$ in the
nanosecond time regime.
The most direct measurements of the RFIM critical
dynamics are of the peak height of the ac susceptibility.  The first
susceptibility measurements on a RFIM antiferromagnet
were performed on $GdAlO_3:La$ by Rohrer~\cite{r81}.  Although
at the time it was thought that the very rounded transition
was evidence of the destruction of the $d=3$ RFIM transition,
it is now understood that the rounding is caused by slow dynamics.
King, et al.~\cite{kmj86} measured the peak height in the ac susceptibility
of $Fe_{0.46}Zn_{0.54}F_2$ as a function of frequency and showed
that the behavior is consistent with either a power law behavior
\begin{equation}
\chi '(\omega ) \sim |t|^{-\alpha}F(\omega |t| ^{\nu z}) \quad ,
\label{chi-power}
\end{equation}
where $z \nu \approx 14$ has an unusually large value,
or with activated dynamics with
\begin{equation}
\chi '(\omega ) \sim |t|^{-\alpha}G(\ln \omega ^\theta) \quad ,
\label{chi-activated}
\end{equation}
where $\theta $ is the violation-of-hyperscaling exponent
(Eq.\ \ref{violation}),
as predicted by Villain~\cite{v85}
and Fisher~\cite{f86}.
Later Nash, et al.~\cite{nkj91} extended the measurements on the same sample
to a very large frequency range of
$5 \times 10^{-3} \le \omega /2 \pi \le 10^5 $ Hz and showed
that activated dynamics are favored by the data with $\theta = 1.05 \pm 0.2$.
This is in accord with the violation of hyperscaling
relation $(d- \theta ) \nu = 2 - \alpha $ using the
measured values of $\nu \approx 1$ and
$\alpha \approx 0$, though it has been suggested
that corrections to scaling should be considered~\cite{pb94}.
However, the picture changed substantially when, recently,
Binek, Kuttler and Kleemann~\cite{bkk95} demonstrated that
in $Fe_{0.47}Zn_{0.53}F_2$ the peak in the ac susceptibility
studied previously is not that of the phase transition itself
but rather is due primarily to the dynamics of Griffiths-like
spatial fluctuations~\cite{d94} above $T_c(H)$.
It was shown that the true critical peak corresponding to the phase
transition is but a small peak that is not resolved at low fields
and was therefore missed in earlier studies, as shown in Fig.\ 4.
The shape of the peak is consistent with the exponent
$\alpha \approx 0$ obtained in other experiments.
High resolution measurements
for a frequency range $3 \times 10^{-1} \le \omega \le 3 \times 10^3$ Hz
again show that the peak is adequately described by either
the power law with an unusually large exponent, $z \nu \approx 14$,
or with activated dynamics.
The critical peak, visible only upon ZFC,
is surprisingly weak, indicating that only a small
portion of the spins are involved in the phase transition. 
This is consistent with the very small peak observed in specific
heat experiments at this concentration~\cite{db89}.
The larger peak has been related~\cite{bkk95} to
Griffiths-like instabilities in the temperature range between
$T_c(H)$ and $T_N$, as discussed in the next section.
Further refinement of the theory for the ac susceptibility
peaks and investigation of other samples, for example
$Fe_{0.93}Zn_{0.07}F_2$, may eventually
settle the question of which dynamic model best fits the
$d=3$ RFIM in dilute antiferromagnets.
The unusual RFIM dynamics have also been observed~\cite{kil90} in
$Fe_{0.7}Mg_{0.3}Cl_2$ using Faraday rotation techniques,
where a symmetric logarithmic
peak is seen with rounding.  A fit of the peak height
to a power law behavior yields $z\nu = 8.3$.

\section{The $d=2$ Destroyed RFIM Transition}

\begin{figure}[t]
\centerline{\hbox{
\psfig{figure=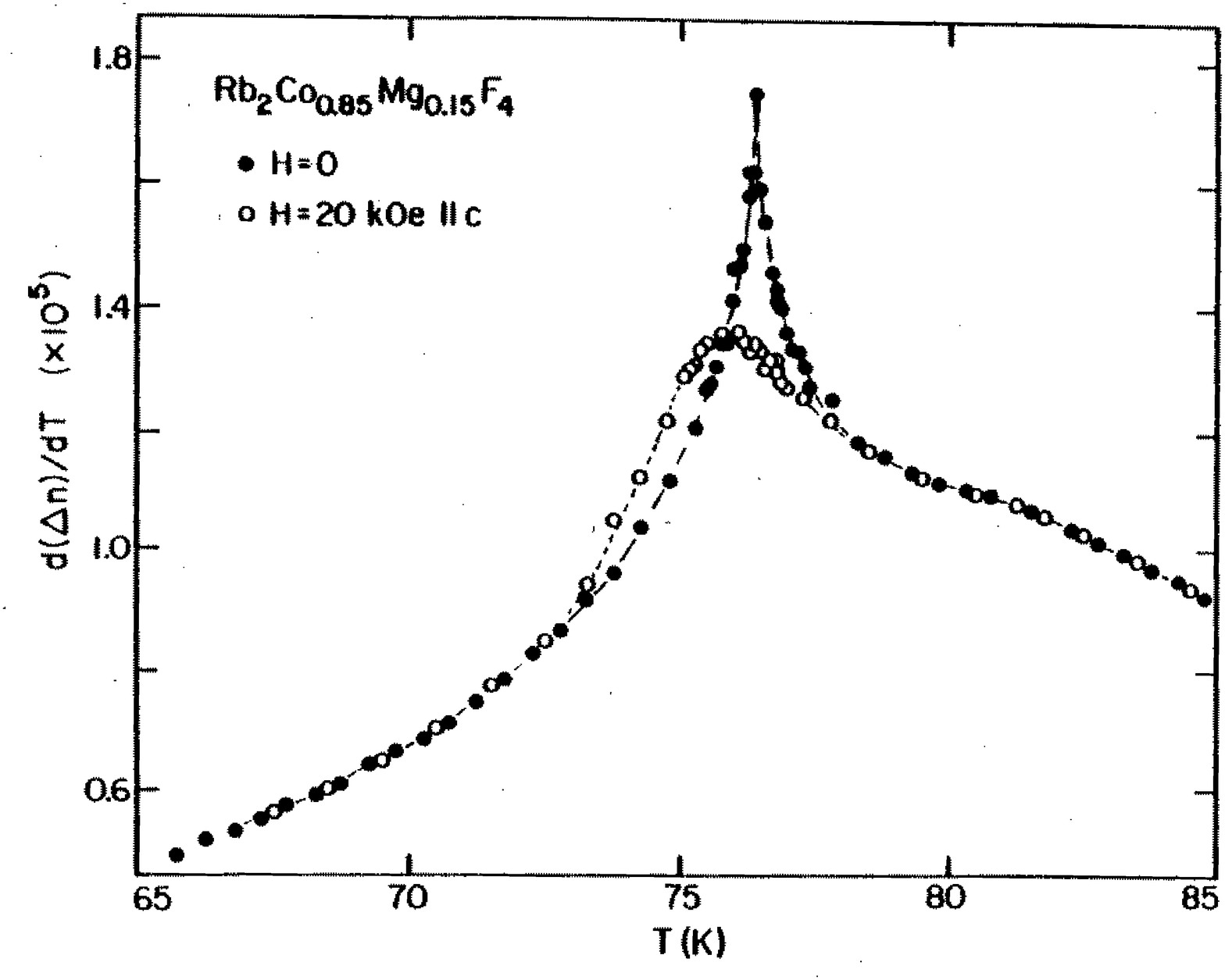,height=2.3in}
\psfig{figure=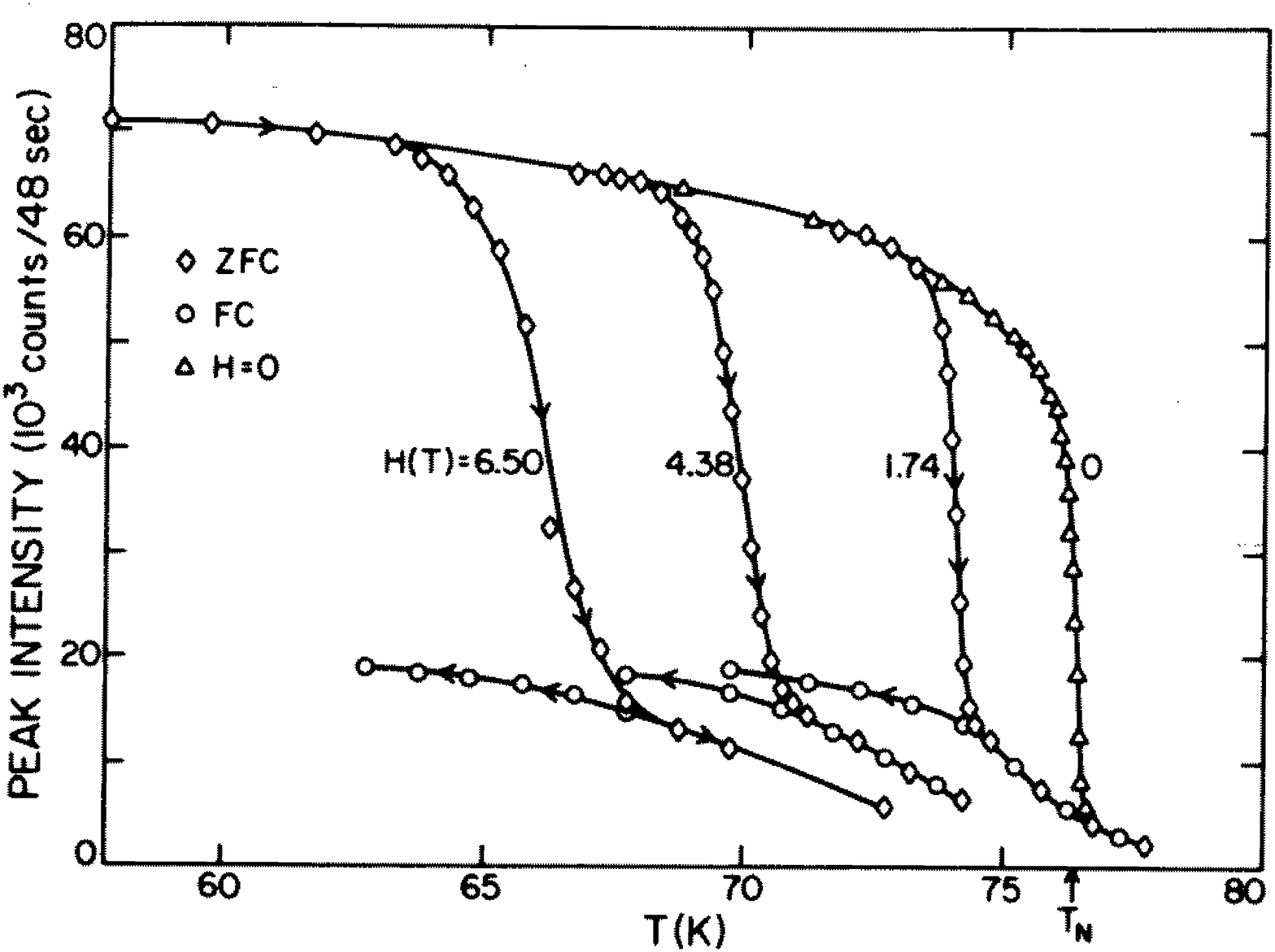,height=2.3in}
}}
\caption{$d(\Delta n)/dT$ vs.\ $T$ and the $q=0$ peak intensity vs. $T$
for the $d=2$ RFIM system $Rb_2Co_{0.85}Mg_{0.15}F_4$.
The birefringence data show
that the application of the random field destroys the transition.  The data
show no hysteresis near $T_c(H)$, which indicates equilibrium behavior.
The neutron scattering peak intensities are obtained after ZFC and FC.
At low temperatures the long-range antiferromagnetic order Bragg component
is stable.
As $T$ is increase, the long-range order becomes unstable and decays,
well below the temperature region of the destroyed phase transition.
No long-range order is observed upon FC.
}
\end{figure}

In contrast to the $d=3$ case, it is clear from theory~\cite{im75,b84}
that the $d=2$ phase transition is destroyed by the random field.
Experimentally this was demonstrated definitively in $Rb_2Co_xMg_{1-x}F_4$
by the birefringence
experiments of Ferreira, et al.~\cite{fkjcg83} and
$(\partial M/ \partial T) _H$ experiments of Ikeda~\cite{i83}.
$\frac {d(\Delta n)}{dT}$ is proportional to the magnetic
specific heat~\cite{fg84} and is particularly important for
low dimensional systems where the phonon specific heat
is considerable.  The transition for $H=0$ is well described
experimentally by a symmetric logarithmic divergence.
However, even relatively small applied fields round the
transition, as is evident in Fig.\ 5.
This behavior contrasts greatly with $d=3$ rounding observed
upon FC since the $d=2$ crystal is in equilibrium above and below the
$H=0$ transition and no hysteresis is observed upon FC and ZFC.
As the field increases, the rounding also increases in
a way consistent with the random-field scaling
function (Eq.\ \ref{scalefunct}) with a crossover exponent $\phi = 1.75$,
which is approximately equal to the zero-field staggered
susceptibility exponent~\cite{hcni87} as expected.

Whereas the behavior near $T_N$ is in excellent accord with theory,
the physics of $d=2$ dilute antiferromagnets at low temperatures
may not be equivalent to that of the ferromagnet with random
fields~\cite{ooq91,dda85}.  The low $T$ behavior
is discussed in the next section.

The neutron scattering line shapes were studied when the sample
was FC to temperatures well below the destroyed phase
transition~\cite{bycsi83} where nonequilibrium behavior dominates.
The Lorentzian plus squared-Lorentzian
line shape of Eq.\ \ref{lor_lor2} works quite well as does a
Lorentzian to a power of approximately $3/2$.
Although no comprehensive study has been 
made of the line shapes near $T_c(H)$, preliminary indications
are that the squared-Lorentzian scattering term is relatively
unimportant in $Rb_2Co_{0.85}Mg_{0.15}F_4$ in this equilibrium
region~\cite{bkj}.  This conflicts with the mean-field theory
that predicts that the squared-Lorentzian should be just as
important for $d=2$ and $d=3$ near $T_c(H)$ and suggests that
the mean-field arguments for the squared-Lorentzian are not particularly
relevant.  This problem deserves further study.

\section{Low Temperature Dynamics in $d=3$ and $d=2$}

The dynamics of the RFIM below $T_c(H)$ in dilute antiferromagnets have
been explored using Squid magnetometry in
$Fe_{0.46}Zn_{0.54}F_2$ by Lederman, et al.~\cite{lsboh93}
After inducing domains using the FC procedure,
the relaxation of the metastable domain walls has been measured as a function
of time.  The excess magnetization from the domain walls scales as the
inverse of the domain size~\cite{lkf88a}.  The dynamics for a variety of
fields and temperatures have been characterized.  Below $T_2(H)$, which is
approximately equal to the equilibrium line $T_{eq}(H)$,
and above the another line $T_1(H)$, the time dependence of the
domain wall size $R(t)$ is consistent with the expression
introduced by Villain~\cite{v84},
$R(t) \sim H^{-\nu _H} \ln (t/\tau )$, where $\tau$ is a spin-flip time.
This indicates that the dynamics are governed by the pinning
from the random-field fluctuations.  Below $T_1(H)$
at lower fields, the random-field pinning seems to be
insignificant relative to the pinning from vacancies,
which are known to freeze in domain structure even
at zero field~\cite{pkb88,nu91,nv88} for $d=3$.
At very low $T$ for all fields the Ising character of the spins
is sufficient to freeze the domain structure.
It is not yet clear how this picture might change with
variation in the magnetic concentration.
The time dependence observed by Lederman, et al.
is consistent with recent domain growth near $T_c(H)$ observed
by Feng, et al.~\cite{fbh95} in $Fe_{0.5}Zn_{0.5}F_2$ in
a very large~\cite{llm96} field $H=5.5T$.  The line shape width decreases
with time near $T_c(H)$ but not at low temperatures.
The smaller field behavior has not been probed.  RFIM dynamics
have also been observed using Monte Carlo techniques~\cite{o94}.

The low $T$ dynamics of the $d=2$ RFIM dilute
antiferromagnet $Rb_2Co_{0.85}Mg_{0.15}F_4$ were probed
using neutron scattering techniques~\cite{bkj85a}.
No Bragg peak develops upon cooling with $H>0$ since
the equilibrium phase transition is destroyed~\cite{fkjcg83}.
Instead, a non-Lorentzian-like scattering line shape
develops~\cite{ycsbgi82} well below the rounded transition.
On the other hand, if the system is ZFC,
long-range order is observed to be stable at low $T$ for $H>0$.  Upon
heating, a temperature region is reached
where the Bragg scattering peak decays.  This region, shown
in Fig.\ 5, is well below the destroyed phase transition as seen by
comparing with the birefringence data~\cite{fkjcg83},
also in Fig.\ 5.  The time dependence of the decay of the Bragg intensity
at the steepest slope in the Bragg intensity, $T_F$,
versus $T$ is observed to be approximately logarithmic.  Furthermore, the
scaling behavior $T_N - T_F \sim H ^ {2/\phi}$
is observed with $\phi = 1.74 \pm 0.02$, in good agreement with
the random-field crossover exponent~\cite{fkjcg83}
$\phi \approx 1.75$.
Hence, the instability of the long-range order is certainly
connected with the random-field behavior.
Just as in the case of $d=3$, once the domains are introduced
into the system below $T_N$ and the field
is turned off, the domains remain for $T<T_N$ even though the
ground state is long-range order.

The dynamics of domain formation at low temperatures have been studied
very close to the percolation threshold in $Rb_2Co_{0.60}Mg_{0.40}F_4$
by Ikeda, et al.~\cite{iei90} using neutron and magnetization techniques.
Currently, the behavior is being investigated~\cite{bjkb96} at higher
concentration in $Rb_2Co_{0.85}Mg_{0.15}F_4$.

\section{Griffiths-like Phase in Dilute Antiferromagnets}

Griffiths~\cite{g69} showed that the magnetization in dilute magnets
is nonanalytic in $H$ at $H=0$ below
the transition temperature of the corresponding pure
system.  This is a consequence of the randomness of
the local magnetic concentration.  Evidence for
dilution-induced Griffiths instabilities has been
observed~\cite{bk95} by studying the deviations from the Curie-Weiss
behavior of $\chi '$ which appears at the pure
N\'{e}el temperature and extends down to the transition
temperature in $Fe_{0.47}Zn_{0.53}F_2$ and $K_2Cu_{0.8}Zn_{0.2}F_2$.
A similar, but much stronger effect is observed in
$Fe_{0.47}Zn_{0.53}F_2$ once random fields are introduced.
Binek and Kleemann~\cite{bk95,bkk95} were able to describe the field-induced
Griffiths-like peak in $\chi '$, seen as the broad peak
in Fig.\ 4,
using a phenomenological Lorentzian density distribution of
local critical temperatures between $T_c(0)=T_N$ and $T_c(H)$ with a
corresponding power law $\chi '$ behavior at each
temperature.  These phenomena have only recently been
investigated~\cite{bjkb96} in $d=2$ systems.

\section{The $d=3$ RFIM at Large Magnetic Dilution and Large Fields}

New physics emerges once the percolation threshold $x \approx 0.24$
in $Fe_xZn_{1-x}F_2$ is approached.  The system behaves much like
a spin-glass~\cite{by86,ymai87}, as was first discovered by Montenegro, et
al.~\cite{mlcr90}$^-$\cite{b81}.
This behavior takes place even though the frustrating exchange interactions
in $Fe_xZn_{1-x}F_2$ are very small~\cite{hrg70}.
Near the percolation threshold, even tiny
frustrating interactions are predicted to become important~\cite{syp79}.
For Ising systems, it is also expected that the dynamics even in
zero field should be extremely slow~\cite{h85}.
Both of these may contribute to the spin-glass-like behavior,
although computer simulations seem to indicate that the small
frustrating interactions are sufficient~\cite{syp79,rcm95}.
Very close to the percolation threshold, for $x=0.25$ and $x=0.27$,
no Bragg peak, and hence no antiferromagnetic ordering, is observed
in zero field
with neutron scattering~\cite{by93}.  (Interestingly, this does not
seem to have been observed in the related anisotropic system~\cite{hcsysbg83}
$Co_{0.26}Zn_{0.74}F_2$ or in the weakly anisotropic system~\cite{csbsg80}
$Mn_xZn_{1-x}F_2$.) The antiferromagnetic
correlation length $\xi$ remains small and constant for $T$ below
approximately $10$~K ($T_N=78.4$ K for pure $FeF_2$).
M\"{o}ssbauer measurements indicate a competition between antiferromagnet
and spin-glass-like order~\cite{acvammrc91}.
The temperature below which $\xi$ remains constant is just the
endpoint of the de Almeida-Thouless line $T_{eq}(H)$.
The $T_c(H)$ curvature is described
well by a crossover exponent $\phi=3.4$, the same exponent
measured in canonical spin-glasses~\cite{by86}.  For a higher concentration,
$x=0.31$, a more complicated phase diagram is
observed~\cite{mlcr90}.  The low-field behavior is the
same as observed for higher 
concentrations, i.e. the low-field phase is antiferromagnetic
and $\phi =1.42$.  As the field increases, the
curvature changes to $\phi=3.4$ and no antiferromagnetic order
is observed below $T_{eq}(H)$.  The large field induces the
spin-glass-like behavior away from percolation.
As we move to even higher concentrations, $x=0.5$, very high fields
are needed to probe the region above the antiferromagnetic phase,
as shown by Lima, et al.~\cite{llm96} employing high-field
magnetization measurements.
Computer simulations~\cite{rcm95} indicate that below
$x \approx 0.6$, weak frustration affects the ordered
state of the REIM in dilute antiferromagnets.

In the less anisotropic system $Mn_{0.35}Zn_{0.65}F_2$,
somewhat similar behavior to that in $Fe_{0.31}Zn_{0.69}F_2$ is
observed~\cite{mjr94} in magnetization
and ac susceptibility studies.  There is some
indication that the phase diagrams may differ in some respects
and this is currently under
investigation.  A de Almeida-Thouless line with $\phi=3.4$ is
observed for $Mn_{0.35}Zn_{0.65}F_2$.

A spin-glass-like phase has also been observed above
the mixed phase in $Fe_xMg_{1-x}Cl_2$ for relatively large
magnetic concentrations~\cite{bfmpk92}.
Slow dynamics are observed for the metastable
domain structure within the mixed antiferromagnetic-paramagnetic
phase~\cite{mkbfk94}.  The memory of domain structure is preserved upon
decreasing the field to zero and even upon field reversal.
The memory effect is also observed after entering the spin-glass-like
phase.

\section{First-order to Second-order Transition in $Fe_xMg_{1-x}Cl_2$}

Recently the metamagnetic transition
in $Fe_xMg_{1-x}Cl_2$ for has been studied
optically and with computer simulations~\cite{kk95}.
Rounding of the metamagnetic transition is interpreted
as the driving of the transition from first-order to
second-order by random fields and random-field-induced domain
structure.  The domain structure is optically observed
to be greatly altered
by the dilution-induced random fields.  This is in accord with
predictions that quenched impurities~\cite{iw79}
and random fields~\cite{sdkkrs93} can drive a phase
transition from first-order to second-order.
The concentration at which the metamagnetic transition
becomes second-order is estimated to be $x=0.6$.
For sufficient dilution the first-order nature of the
transition is lost when the avalanche of domain flipping no longer
involves infinite length scales.
Universal behavior is predicted for this nonequilibrium
transition~\cite{ds96}.

\section{Other RFIM Systems}

Although a great deal of the experiments shedding light on the
RFIM have been done on dilute antiferromagnets, other systems
have been studied as well.  Kleemann~\cite{k93} has reviewed
random-field domain states in ferroelectric and structural phase
transitions.  The critical behavior of the RFIM structural phase transition
in $DyAs_xV_{1-x}O_4$
has been studied extensively~\cite{sbwt96} and compared to the
dilute antiferromagnet.  Neutron and light scattering  experiments
have been done on binary mixtures in silica gels~\cite{fcls95}.
Certainly more RFIM realizations will be studied in the future and will
significantly add to our understanding as well as incorporate aspects
of the dilute antiferromagnet results.

\section{Conclusions}

There is good reason to be optimistic about achieving
a good characterization of $d=3$ RFIM critical behavior
in the near future.  Experiments
are nearly at the point where serious comparisons between theory and experiment
can be made.  This is possible since the high concentration
crystals show no evidence for the formation of domain
structure or for hysteresis in the line shapes well below the transition,
two aspects of the experiments at lower concentration that have
been severe impediments.  It would be interesting to
investigate if the remarkable difference in the
behavior at high and low magnetic concentration is
a result of a concentration critical point below which the
long-range order becomes unstable.
One outstanding problem is the lack of a theoretically derived
line shape to use in analyzing data; the mean-field arguments
are clearly inadequate.  When such a theory is developed, more reliable
critical exponents and amplitude ratios will be derived from
the scattering data.

Two kinds of hysteresis can now be distinguished for $d=3$.
At low magnetic concentrations, vacancies cause irreversibilities
and domain formation which are most evident in scattering experiments.
At all concentrations where transitions
take place, there appears to be hysteresis, observable in all
experiments, that may be attributable to random-field critical dynamics.
For $d=2$, hysteresis occurs only at low temperatures, well below
the rounded transition.  The dynamics of domain formation in this
region are still being studied.

Griffiths-like domain structure dominates the ac suseptibility
in the $d=3$ random-field transition in $Fe_{0.46}Zn_{0.54}F_2$.
It remains a task to determine whether the small critical
peak, recently discovered, yields power-law or activated dynamics.

Near the percolation threshold, it appears that the
$d=3$ Ising system $Fe_xZn_{1-x}F_2$ behaves very much like
a spin-glass despite having only small frustrating interactions.
The behavior in the more isotropic $Mn_xZn_{1-x}F_2$ is being
studied to elucidate the role of anisotropy in the spin-glass-like
behavior.  In related studies, intermediate concentration
crystals of $Fe_xZn_{1-x}F_2$ are being studied in the high-field limit.

Recent experiments have addressed the random-field
effects on first-order transitions in $Fe_xMg_{1-x}Cl_2$.
The first-order transition appears to be driven to be second-order
with sufficiently strong random fields, in agreement with theory.

A reasonable understanding of the random-field Ising model
as realized in dilute antiferromagnets is emerging, though
there is considerable work yet to be done.
All aspects of the rich behavior of these dilute antiferromagnets
are important to characterize partly for their intrinsically
interesting properties and partly because other materials may
show one or more of the characteristics.  The antiferromagnets
are the best studied and probably the most easily understood
systems.  In trying to understand the behavior in more
complex systems, one will have to keep in mind the array
of possible behaviors.  Certainly, the random-field physics
will be incorporated into the descriptions of many important
materials in the future.

\section*{Acknowledgements}

Some of the experiments on the $d=3$ $Fe_xZn_{1-x}F_2$
would not be possible without the extraordinary crystals
produced by Vince Jaccarino, Allan King and Neil Nighman.  I thank
Vince Jaccarino for giving some of them to me.
I would like to thank Zoran Slani\v{c}, Jaime Fernandez-Baca, Wolfgang
Kleemann, Frederico Montenegro, Christian Binek and others for
allowing me to review their data prior to publication and
for interesting discussions.  In addition to those above, I would like to
acknowledge Bob Nicklow, Hideki Yoshizawa,
Jiahua Wang, Seung-Jin Han, Keith Dow, Carlos Ramos, David Lederman,
Mark Lui, Sergio Rezende, Peter Pollak, Ingrid Ferreira, and others with whom I enjoyed collaborating on some of the work
reviewed here.  I would like to thank Peter Young, Uli Nowak and others
for interesting and helpful discussions.  This work has been supported
by Department of Energy Grant No. DE-FG03-87ER45324.


\begin{thebibliography}{999}
\bibitem{fl}It has recently been claimed~\cite{bfhhrt95}
that justification
for seeing the antiferromagnetic order parameter in
magnetization or birefringence measurements can be found
in the work of Langer and Fisher~\cite{fl68}.
However, that reference is the basis of seeing the
critical behavior of the specific heat in resistivity
measurements and the argument can be developed further~\cite{fg84}
to show why birefringence exhibits the
specific heat critical behavior.  It does not
address long-range antiferromagnetic order.
\end{thebibliography}
\end{document}